\documentclass[9pt,twocolumn,twoside]{osajnl}

\usepackage{amsmath}
\usepackage{amssymb}

\journal{ol} 

\setboolean{shortarticle}{true}

\title{Robust stability of quantum interference realized by coexisting detuned and resonant STIRAPs}

\author[]{Yichun Gao}
\author[]{Jianqin Xu}
\author[*]{Jing Qian}

\affil[]{State Key Laboratory of Precision Spectroscopy, Quantum Institute for Light and Atoms, Department of Physics, School of Physics and Electronic Science, East China Normal University, Shanghai 200062, China}

\affil[*]{Corresponding author: jqian1982@gmail.com}

\begin{abstract}
Inspired by a recent experiment [Phys. Rev. Letts. \textbf{122}, 253201(2019)] that an unprecedented quantum interference was observed in the way of Stimulated Raman adiabatic passage (STIRAP) due to the coexisting resonant- and detuned-STIRAPs, we comprehensively study this effect for uncovering its robustness towards the external-field fluctuations of laser noise, imperfect resonance condition as well as the excited-state decaying. We verify that, an auxiliary dynamical phase accumulated in hold time caused by the quasi-dark state can sensitively manipulate the visibility and frequency of the interference fringe, representing a new hallmark to measure the hyperfine energy accurately. The robust stability of scheme comes from the intrinsic superiority embedded in STIRAP itself, which promises a remarkable preservation of the quantum interference quality in a practical implementation.
\end{abstract}

\setboolean{displaycopyright}{true}

\begin{document}

\maketitle

Quantum interference effect (QIE) serving as one of the most intriguing features that differs a quantum system from classical candidates, has facilitated versatile applications in diverse systems, covering the range from electron source \cite{Bauerle:18}, single-atom-cavity system \cite{Tang:19}, superconducting device \cite{Lecocq:17} to solid-spin system \cite{Bernien:12}. For realizing QIE, all the accessible routes of different platforms essentially require at least more than two possible channels for transferring the information, returning back a representation of small forces or energies in precision measurement \cite{Muller:10}, quantum entanglement \cite{Ott:10} or quantum sensing \cite{Degen:17} at the microscopic level. In reality, QIE based on the quality of a quantum system is still very limited due to the imperfect stability of external magnetic or optical fields under real implementation, which can sensitively dephase the interference by artificial measurement or noise effect \cite{Comparat:20}. Although a large number of approaches for overcoming the instability or imperfect measurement have been proposed, {\it e.g.} quantum nondemolition measurement can enable a repeated detection of quantum states without destroying it \cite{Braginsky:96}; quantum plasmonics experiments exhibit a remarkable preservation of coherence \cite{Tokpanov:19}, a more precise atom interferometry is achieved by transferring the photon momentum to atoms while minimizing its uncertainty \cite{McAlpine:20}; it is still challenging for realizing an extremely stable QIE in well-isolated systems.

Thanks to the contributions by Liu and coworkers that an unprecedented observation of QIE induced by the Stimulated Raman adiabatic passage (STIRAP) was achieved recently \cite{Liu:19}, arising a new avenue to precise measurement utilizing an absolute adiabatic system \cite{Weitz:94}. STIRAP technique basically benefiting from a well coherence-preservation, has been widely used for a determined population transfer among ground states with high efficiency \cite{Vitanov:17}. Unfortunately only (two-photon)resonant-STIRAP(R-STIRAP) is favored by previous studies owing to a complete dark-state isolated from the influence of other bright states. While the effect of detuned-STIRAP(D-STIRAP) is usually ignored \cite{Deng:14} because of the robustness of coexisting R-STIRAP making the impact of D-STIRAP very weak. The realization of QIE by STIRAPs unexpectedly reveals the virtue of D-STIRAP even if the R-STIRAP is active, which can provide an accurate measurement for the small hyperfine energy in ground-state levels.

Inspired by their experimental facts \cite{Liu:19}, in the present work, we comprehensively study the intriguing feature of QIE induced by coexisting R-STIRAP and D-STIRAP in a four-level $\Lambda$ system, uncovering the importance of D-STIRAP by adding a controllable dynamical phase between the two quantum paths. The interference frequency and visibility that quantitatively rely on the strength of two-photon detuning in D-STIRAP, can be used as a reliable measurement tool for small energy differences in a practical implementation. Furthermore the observed interference fringe can keep a stable output persistently against significant stochastic fluctuations from the laser noises as well as the limited lifetime of excited-states, which are mainly favored by a remarkable stability possessed by an adiabatic STIRAP system itself \cite{Bergmann:98}.


\begin{figure}[htbp]
\centering
\fbox{\includegraphics[width=\linewidth]{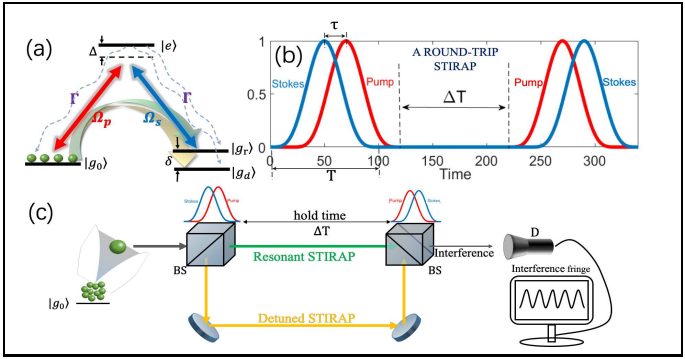}}
\caption{(a) The four-level $\Lambda$ structure and the atom-field interaction implemented by a round-trip STIRAP pulses. R-STIRAP and D-STIRAP are denoted by green and yellow arrows. (b) The time diagram of STIRAP pulses in which $\Delta T$ stands for the hold time interval between two pulse pairs. (c) The proof-of-principle experiment for realizing a STIRAP atom interferometry.}
\label{fig1}
\end{figure}

As represented in Fig.\ref{fig1}(a), each atom consists of the least four levels as denoted by $|g_0\rangle$, $|e\rangle$, $|g_{r,d}\rangle$ for describing the initial ground, middle excited and hyperfine ground states. Two pairs of STIRAP sequences are introduced with Rabi frequencies $\Omega_p$ and $\Omega_s$ to make resonant transitions between $|g_0\rangle\leftrightarrow|e\rangle$ and $|e\rangle\leftrightarrow|g_r\rangle$, forming a round-trip R-STIRAP channel. At the same time the adjacent hyperfine ground state $|g_d\rangle$ detuned by $\delta$ with respect to $|g_r\rangle$ presents a D-STIRAP of $|g_0\rangle\rightleftarrows|e\rangle\rightleftarrows|g_d\rangle$ with same pulses . The population undergoing R- and D-STIRAPs will interfere leading to a population oscillation on state $|g_0\rangle$ versus the hold time $\Delta T$. That can be served as a way for stabilizing atom interferometry.
In the rotating-wave frame, the Hamiltonian given by $\hat{\mathcal{H}}=-(\Delta\hat{\sigma}_{ee}+\delta\hat{\sigma}_{g_dg_d})+\frac{1}{2}[\Omega_p\hat{\sigma}_{g_0e}+\Omega_s(\hat{\sigma}_{eg_r}+\hat{\sigma}_{eg_d})+H.c.](\hbar=1)$ explains the atom-light interactions where operator is 
$\hat{\sigma}_{ij}=|i\rangle\langle j|$ and $\Delta$ (or $\delta$) is treated as an one(or two)-photon detuning. It is obvious that $\delta$ also refers to the hyperfine energy difference of two ground states, typically scaling of $\sim kHz$ \cite{Das:08}. The STIRAP pulse as in Fig. \ref{fig1}(b) takes a generalized form of 
\begin{equation}
\begin{aligned}
&\Omega_p(t)=A(t-\tau)+A(t-\tau-\Delta T-T),\\
&\Omega_s(t)=A(t)+A(t-2\tau-\Delta T-T),
\end{aligned}
\label{eqlight}
\end{equation}
where the amplitude function $A(t)$ is given by
\begin{equation}
A(t)\equiv
\left\{\begin{aligned}
        &\Omega_{0,p(s)}sin^4(\frac{\pi t}{T}) &   0\leqslant t \leqslant T\\
        &0 &   otherwise.
       \end{aligned}
\ \right.
\end{equation}
and $T$ is the common pulse length, $\Omega_{0,p(s)}$ are for the peak amplitudes of {\it pump}({\it Stokes}) lasers. For simplicity $\Omega_0 = \Omega_{0,p(s)}$ is assumed.

\begin{figure}[htbp]
\centering
\fbox{\includegraphics[width=\linewidth]{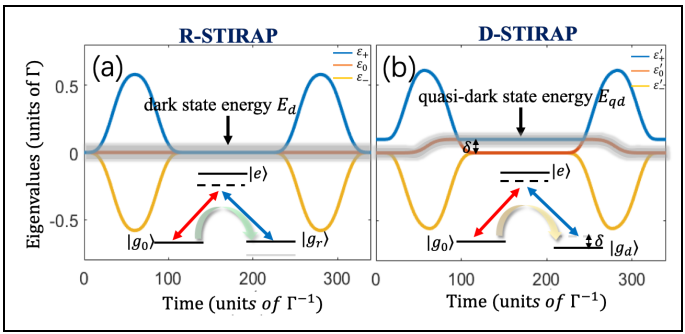}}
\caption{(a) In a reduced three-level R-STIRAP structure the eigenvalues as a function of time. The dark-state energy is $E_d=0$ as denoted by the grey shaded area. Similar to (a) the quasi-dark energy $E_{qd}$ in the case of D-STIRAP is given in (b).}
\label{fig2}
\end{figure}

\begin{figure}[htbp]
\centering
\fbox{\includegraphics[width=\linewidth]{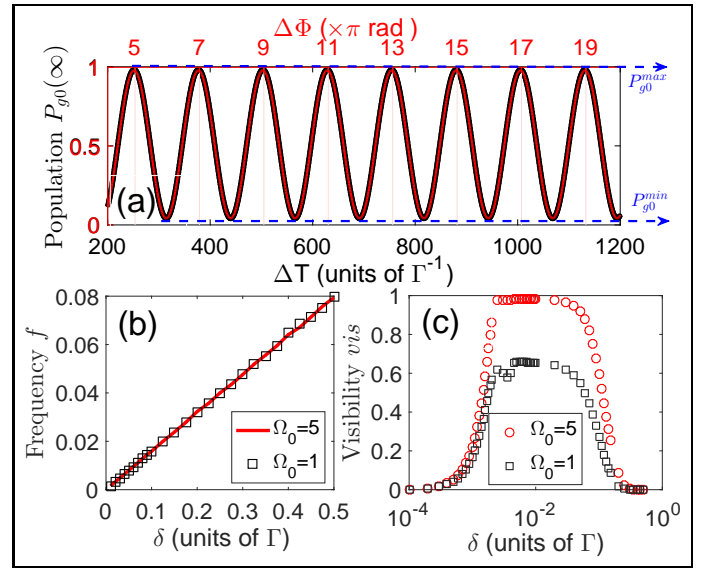}}
\caption{(a) The population of state $|g_0\rangle$ denoted by $P_{g0}(\infty)$ oscillates after a round-trip STIRAP, as a function of the hold time $\Delta T\in[200,1200]/\Gamma$ for $\delta=0.05$, $\Omega_{0}=5.0$. A parallel axis denoted by the accumulated phase $\Delta\Phi$ is also presented. The frequency (b) and visibility (c) of the interference fringe are comparably presented with respect to the hyperfine energy difference $\delta$. Here $\Omega_{0}=5.0$(red), $1.0$ (black) and  $\Gamma$($\Gamma^{-1}$) treats as the frequency(time) unit.}
\label{fig3}
\end{figure}

The proof-of-principle formation of QIE can be traced to a combination of R-STIRAP along $|g_0\rangle\leftrightarrows|e\rangle\leftrightarrows|g_r\rangle$ as well as D-STIRAP along $|g_0\rangle\leftrightarrows|e\rangle\leftrightarrows|g_d\rangle$ in reduced three-level structures. By ignoring the one-photon detuning $\Delta$ it is easier to solve all eigenvalues in reduced schemes, which are $E_d=\varepsilon_0=0$, $\varepsilon_{\pm}=\pm\sqrt{{\Omega_p}^2+{\Omega_s}^2}/2$, and ${\varepsilon^{'}}_{0}=(\delta+\widetilde{\Omega}\cos\frac{\zeta}{3})/3$, ${\varepsilon^{'}}_{\pm}=(\delta+\widetilde{\Omega}\cos\frac{2\pi\mp\zeta}{3})/3$ with $\zeta=2\pi-\arccos{\frac{[9({\Omega_p}^2+{\Omega_s}^2)+8{\delta}^2]\delta-27{\Omega_p}^2\delta}{{\widetilde{\Omega}}^3}}$ and $\widetilde{\Omega}=\sqrt{3({\Omega_p}^2+{\Omega_s}^2)+4{\delta}^2}$, as displayed in Fig.\ref{fig2}(a) and (b).
The composite quasi-dark energy $E_{qd}$ (shaded in grey in (b)) differs from $E_d$ by a value $\delta$ during the hold time when $\Omega_p = \Omega_s = 0$, which enables an accumulated dynamical phase $\Delta\Phi$ as the population evolves in D-STIRAP. A numerical study for this phase $\Delta\Phi$ affecting on the interference fringe can be carried out by solving the master equation $\frac{d\hat{\rho}}{dt}=i[\hat{\rho},\hat{\mathcal{H}(t)}]+\hat{\mathcal{L}}(\hat{\rho})$ with Lindbald operator given by $\hat{\mathcal{L}}(\hat{\rho})=\Gamma\sum_{j\in\{g_0,g_r,g_d\}}[\hat{\sigma}_{je}\hat{\rho}\hat{\sigma}_{ej}-\frac{1}{2}(\hat{\sigma}_{ee}\hat{\rho}+\hat{\rho}\hat{\sigma}_{ee})$. Note that $\Gamma$ is the spontaneous decay rate from the excited state $|e\rangle$ limited by its lifetime, and $P_{g0}= \rho_{g_0g_0}(t)$ serves as the main observed quantity in detection. As predicted by experiment \cite{Liu:19} we also find a strong oscillation of $P_{g0}(\infty)$ versus the change of $\Delta T$, see Fig. \ref{fig3}(a) at $\delta/\Gamma = 0.05$.

 The cause of oscillation could be qualitatively understood with the frame of STIRAP atom interferometry as in Fig. \ref{fig1}(c). Starting from the case of all population on state $|g_0\rangle$, the first counterintuitive pulses ($\Omega_s$ exceeds $\Omega_p$) act as a "beam splitter"(BS) that coherently splits into two STIRAP channels for the population conversion, giving rise to a final superposition state containing the populations on $|g_r\rangle$ and $|g_d\rangle$ after the first pulse pair. However, before the arrival of second inverse pulses ($\Omega_p$ exceeds $\Omega_s$) there exists a hold time $\Delta T$ permitting a free evolution of these populations following the dark- or quasi-dark states, accordingly. To this end the population of $|g_d\rangle$ would acquire a relative dynamical phase $\Delta\Phi$ due to its shifted energy $\delta$, analogous to the undergoing of a different optical path. The interference fringe after the second BS (reverse pulses) could be precisely determined by this relative phase. A quantitative estimation for the phase can be expressed as $\Delta\Phi=\Phi_d-\Phi_r = \varphi_c+\delta(\Delta T) $ with $\Phi_d=\int_0^t E_{qd}(t^{'}) dt^{'}=\varphi_c+\delta(\Delta T)$, $\Phi_r=\int_0^t E_d(t^{'}) dt^{'}=0$ for the exact accumulated phase in D- and R-STIRAPs, where $t=T+\tau+\Delta T$ is the total evolutionary time before the reverse STIRAP. $\varphi_c$ is a constant phase acquired during the forward STIRAP. Hence the phase difference $\Delta\Phi$ is mainly determined by $\delta(\Delta T)$, leading to the frequency $f=(1/2\pi)\delta$ versus $\Delta T$.

 A numerical identification for the relationship between the frequency $f$ or visibility $vis$ of interference fringe and $\delta$ is presented in Fig.\ref{fig3}(b-c). As expected
 the linear relation of $f$ and $\delta$ is absolute that does not depend on the peak amplitudes $\Omega_{0}$. However, the interference contrast ({\it i.e.} visibility) defined by: $vis = (P_{g0}^{max} - P_{g0}^{min})/(P_{g0}^{max} +P_{g0}^{min})$ is sensitive to $\delta$ due to the effective competing effect between R- and D-STIRAPs. As far as we know, as $\delta\to0$ the two STIRAPs can converge into one resonant adiabatic passage causing a big fall of visibility. As increasing $\delta$, {\it vis} will enter its maximum at $\delta(\Delta T)=2\pi$ and $\Delta T\in[200,1200]/\Gamma$ for constructive interference. While in the opposite case if $\delta$ is significantly increased, even to be a same level as the Rabi field, a big detuning from resonance will make R-STIRAP invalid in the competition, arising a poor visibility yet. It is stressed that a suitable range of $\delta$ is required for realizing an optimal $vis$ that does not depend on the exact laser Rabi frequencies. Also $vis$ can be deeply increased even closing to unity at $\Omega_{0}=5.0$, since a smaller $\Omega_{0}$ may break the adiabaticity $r$ of STIRAP due to $r\propto\Omega_{0}^{-1}$ \cite{Pu:07}, suffering from an inadequate population transfer. Therefore it is reliable to determine the hyperfine splitting energy $\delta$ by detecting the oscillating frequency $f$, interactively returning a feedback of precise measurement for the ground-energy under a practical application.

\begin{figure}[htbp]
\centering
\fbox{\includegraphics[width=\linewidth]{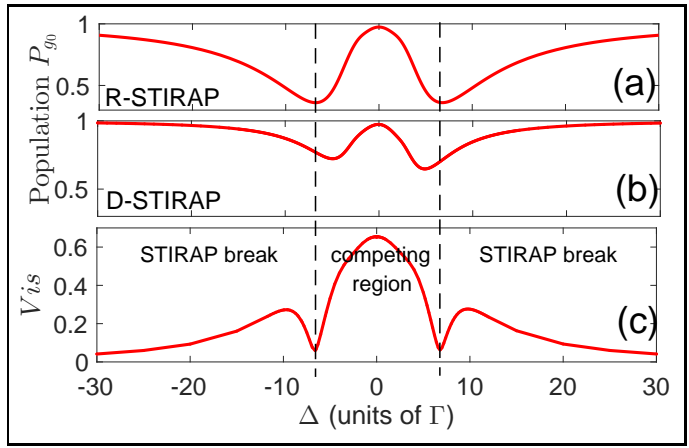}}
\caption{(a-b) Representation of the final population of $|g_0\rangle$ as a function of the intermediate detuning $\Delta$. (c) The visibility of interference fringe labeled by {\it vis} coming from the population interferes on state $|P_{g0}\rangle$, with respect to the detuning $\Delta$. Here $\Omega_{0} = 1.0$, $\delta = 0.05$. }
\label{fig4}
\end{figure}

The competing effect of R- and D-STIRAP can also be manifested by a variable one-photon detuning $\Delta$. Notice that the energy difference between adjacent excited hyperfine levels with respect to $|e\rangle$ typically $\sim100$MHz, is orders of magnitude larger than the laser Rabi frequencies, arising a safe negligence of other excited states by retaining the validity of four-level protocol. However if this unique excited state $|e\rangle$ is shifted by $\Delta$, it may also lead to a reduction of transfer efficiency, resulting in a poor fringe contrast. Studying this effect can uncover the competing region where both STIRAPs play roles. As increasing $|\Delta|$ from resonance we detect population $P_{g0}(\infty)$ after all pulses and find that $P_{g0}(\infty)$ can individually exhibit a monotonous decreasing at the early stage, then counterintuitively increases for a larger $\Delta$. That fact also leads to a discontinuity in the variation of {\it vis} versus the increase of $|\Delta|$, as represented in Fig.\ref{fig4}(c). For understanding, if $|\Delta|$ is near-resonance, a coherent population transfer based on two STIRAPs carrying out at the same time, will arise a competing region where an enhanced one-photon detuning would significantly lower the transfer efficiency. However as $\Delta$ exceeds a critical value as pointed by dashed lines, $P_{g0}(\infty)$ is observed to gradually increase accompanied by the entire breakdown of visibility. Because if $\Delta$ is sufficient most population remains in the initial state $|g_0\rangle$ without transmission through the round-trip process, resulting in no QIE. 
To this end, it is remarkable that $\Delta$ as a control knob for adjusting the relative strength between R- and D-STIRAP, can precisely control the performance of STIRAP interferometry, revealing the competing region where both of them definitely react. A well-resolved excited state can be identified if the one-photon detuning is one order of magnitude smaller compared to the Rabi frequencies {\it i.e.} $\Delta/\Omega_{0}\leq 0.1$, as confirmed by experiment.

\begin{figure}[htbp]
\centering
\fbox{\includegraphics[width=\linewidth]{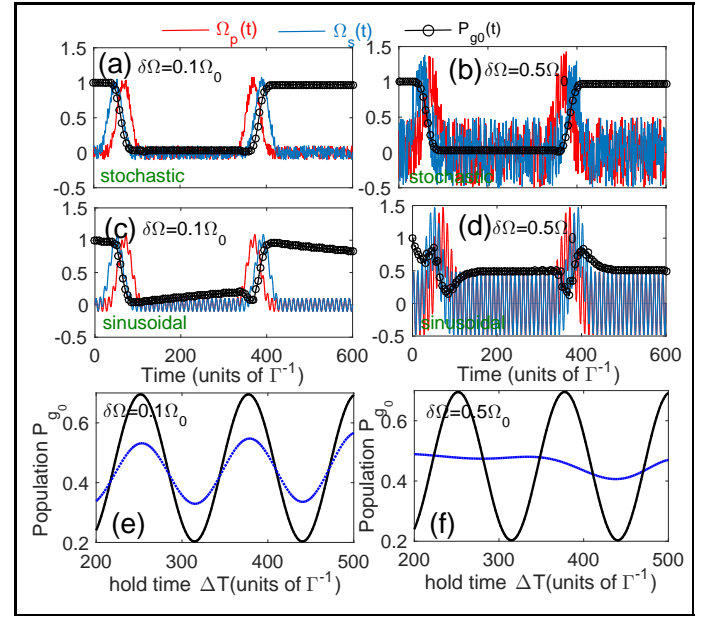}}
\caption{(a-b) The laser pulses $\Omega_p(t)$(red) and $\Omega_s(t)$(blue) under the stochastic fluctuations within the region of $\delta\Omega_{p(s)}(t)\in[-\delta\Omega,\delta\Omega]$. Here $\delta\Omega = 0.1\Omega_0$ for (a) and $\delta\Omega = 0.5\Omega_0$ for (b), accompanied by a detection for the time-dependent population $P_{g0}(t)$(black with circles) on state $|g_0\rangle$; (c-d) are same as (a-b) except for a sinusoidal-type fluctuation. (e-f) The interference fringe from stochastic (black-solid) and sinusoial (blue-dotted) fluctuations, characterized by the population $P_{g0}(\infty)$ oscillating as tuning the hold time $\Delta T$. Here $\Omega_0 =1.0 $. }
\label{fig5}
\end{figure}

As far as we know the use of stabilized lasers is important for a continuous measurement feedback in a matter-wave interferometry \cite{Olson:19}. Because due to the requirement of high measurement accuracy, any shifting of fringe may return back necessary information about {\it e.g.} hyperfine splitting energy, additional optical phase {\it etc.}, quantitatively determining the measurement quality. However the laser system in experiment can hardly acquire absolute stability because of the intensity noise even with the use of laser-frequency stabilization technique, which possibly destroys stable interference output. This noise may lead to a stochastic fluctuation to the laser Rabi frequency, or arising an imperfect two-photon resonance. Benefiting from the intrinsic stability of STIRAP technique that does not depend on absolute laser shapes, our QIE can manifest a robust stabilization towards any stochastic fluctuations from the laser intensities.

To verify that, we first add significant fluctuations to the peak intensity by $\Omega_{s(p)}^{\prime}(t) = \Omega_{s(p)}(t)+\delta\Omega_{s(p)}(t) $ with $\delta\Omega_{s(p)}$ stochastically obtained from $[-\delta\Omega,\delta\Omega]$ in Fig.\ref{fig5}(a-b), and $\delta\Omega_{s(p)}=\delta\Omega\sin(\omega t)$(here $\omega/2\pi = 0.1\Gamma$) in Fig.\ref{fig5}(c-d). It is evident that the population evolution $P_{g0}(t)$(black with circles) enables an adiabatic preservation against significant stochastic perturbations, but easily breaking with a sinusoidal and non-practical adjustment. The reason for that can be understood by the phase accumulation during the hold time $\Delta T$ which critically depends on the quasi-dark energy $E_{qd}$. If $\delta\Omega_{s(p)}=0$ the phase difference is only determined by $\delta(\Delta T)$ revealing a precise relation with the splitting $\delta$. However as $\delta\Omega_{s(p)}\neq0$ it leads to the extra perturbed phase. Thanks to a stochastic fluctuation that arises a complete  cancellation to the phase change on average, persisting the usual phase difference $\Delta\Phi$. But a regular sinusoidal modulation only causes a partial cancellation, resulting in a poor population transfer. Therefore as expected, our protocol can manifest a robust stability towards arbitrary stochastic fluctuations from the peak laser intensity, preserving a high-contrast interference output. In comparison, the interference fringe affected by a sinusoidal modulation is poor which rapidly decreases with $\delta\Omega_{s(p)}$.

\begin{figure}[htbp]
\centering
\fbox{\includegraphics[width=\linewidth]{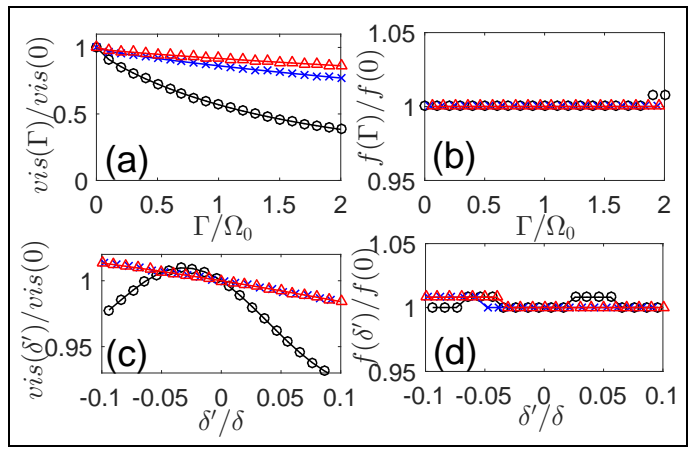}}
\caption{The relative strength of $vis$ and frequency $f$ versus the change of (a-b) increasing the excited-state decay $\Gamma$ and (c-d) a tunable two-photon detuning $\delta^{\prime}$, with respect to the cases of $\Gamma=0$ and $\delta^{\prime}=0$, respectively. The curves with red triangles, blue crossings and black circles are using $\Omega_{0}=$ 10, 5, 1 respectively. }
\label{fig6}
\end{figure}

Second, the scheme robustness can also be shown if the excited-state decay $\Gamma$ is significant or under an imperfect two-photon resonance characterized by $\delta^{\prime}\neq0$. Results have been confirmed in Fig.\ref{fig6} where the oscillating frequency $f$ of interference pattern can persist a perfect invariance no matter how $\Gamma$ or $\delta^{\prime}$ changes due to $f = (1/2\pi)\delta$ at a given $\Delta T$. On the other hand, the visibility $vis$ is indeed affected by the absolute values of $\Gamma$ and $|\delta^{\prime}|$ but enabling a well preservation with the increase of the peak amplitude of Rabi frequency $\Omega_0$. An estimation based on a practical system of $^{87}$Rb atoms can rely on the typically energy levels: $|g_0\rangle = |5s_{1/2},F=1,m_F=-1\rangle$, $|g_{r}\rangle =|5s_{1/2},F=2,m_F=1\rangle$, $|g_d\rangle = |5s_{1/2},F=2,m_F=0 \rangle$, $|e\rangle =|5p_{3/2}\rangle $, which confirms that, at $\Gamma/2\pi = 6.0$MHz, $\Omega_{0}/2\pi = 42$MHz, $T = 2.65\mu$s, $\Delta T\in(5.3\sim31.8)\mu$s, $\delta = 264k$Hz, $vis$ can attain as high as 0.9901 with the oscillating frequency $f=42.04k$Hz.

To conclude the high-quality of QIE can arise a more precise quantum measurement for facilitating an accurate control of hyperfine structure or phase sensitivity. Guided by recent experimental facts that we investigate the formation of a high-contrast QIE with robust stability, which essentially benefits from the intrinsic advantage of STIRAP technology. First the scheme exhibits clear interference fringe with a high visibility, supporting a quantitative and non-destructive detection of hyperfine energy without destroying the coherence of quantum states. Second, the production of QIE has a well-preserved stabilization against any stochastic fluctuations coming from the laser intensity noise, the small shift of two-photon resonance as well as the limited lifetime of middle excited state. These facts are intrinsically enabled by a stabilized STIRAP transfer which is greatly isolated from external influences. This protocol promises an important step towards the development of a stabilized STIRAP atom interferometry with ultrahigh precision for future studies.

\medskip

\noindent\textbf{Funding.} The National Natural Science Foundation of China (11474094, 11104076); The Science and Technology Commission of Shanghai Municipality (18ZR1412800).

\medskip

\noindent\textbf{Disclosures.} The authors declare no conflicts of interest.


\bibliography{references}

\bibliographyfullrefs{references}
  

\ifthenelse{\equal{\journalref}{aop}}{%
\section*{Author Biographies}
\begingroup
\setlength\intextsep{0pt}
\begin{minipage}[t][6.3cm][t]{1.0\textwidth} 
  \begin{wrapfigure}{L}{0.25\textwidth}
    \includegraphics[width=0.25\textwidth]{john_smith.eps}
  \end{wrapfigure}
  \noindent
  {\bfseries John Smith} received his BSc (Mathematics) in 2000 from The University of Maryland. His research interests include lasers and optics.
\end{minipage}
\begin{minipage}{1.0\textwidth}
  \begin{wrapfigure}{L}{0.25\textwidth}
    \includegraphics[width=0.25\textwidth]{alice_smith.eps}
  \end{wrapfigure}
  \noindent
  {\bfseries Alice Smith} also received her BSc (Mathematics) in 2000 from The University of Maryland. Her research interests also include lasers and optics.
\end{minipage}
\endgroup
}{}

\end{document}